\documentclass[aps,showpacs]{revtex4}
\usepackage{graphicx}

\begin{document}
\title{Binary $N$-Step Markov Chain as an Exactly Solvable Model
of Long-Range Correlated Systems}

\author{O.V. Usatenko and V.A. Yampol'skii}
\address{A.Ya. Usikov Institute for Radiophysics and Electronics \\
Ukrainian Academy of Science, 12 Proskura Street, 61085 Kharkov,
Ukraine}

\begin{abstract}
A theory of systems with long-range correlations based on the
consideration of \textit{binary N-step Markov chains} is
developed. In our model, the conditional probability that the
$i$-th symbol in the chain equals zero (or unity) is a linear
function of the number of unities among the preceding $N$ symbols.
The model allows exact analytical treatment. The correlation and
distribution functions as well as the variance of number of
symbols in the words of arbitrary length $L$ are obtained
analytically and numerically. A self-similarity of the studied
stochastic process is revealed and the similarity transformation
of the chain parameters is presented. The diffusion equation
governing the distribution function of the $L$-words is explored.
If the persistent correlations are not extremely strong, the
distribution function is shown to be the Gaussian with the
variance being nonlinearly dependent on $L$. The applicability of
the developed theory to the coarse-grained written and DNA texts
is discussed.
\end{abstract}

\date{\today}


\maketitle

\section{Introduction}

The problem of long-range correlations is one of the topics of
current research in the statistical mechanics. The stochastic
processes with strong correlations have been observed in numerous
systems. Examples include the natural languages, the DNA
sequences, etc.~\cite{stan,prov}.

One of the efficient methods to investigate the correlated systems
is based on a decomposition of the space of states into a finite
number of parts labelled by definite symbols. This procedure
referred to as coarse graining is accompanied by the loss of
short-range memory between symbols but does not affect and does
not damage the robust invariant statistical properties of the
long-range correlated sequences. In terms of the power spectrum,
one loses only the short-wave part of the spectrum. The most
frequently used method of the decomposition is based on the
introduction of two parts of the phase space. In other words, it
consists in mapping the two parts of states onto two symbols, say
0 and 1. Thus, the problem is reduced to the investigation into
the statistical properties of the binary sequences. This method is
applicable for the examination of the both discrete and continuous
systems.

One of the ways to get a correct insight into the nature of
correlations in a system consists in an ability of constructing a
mathematical object (for example, a correlated sequence of
symbols) possessing the same statistical properties as the initial
system. There are many algorithms to generate long-range
correlated sequences: the inverse Fourier transform~\cite{czir},
the expansion-modification Li method~\cite{li}, the Voss procedure
of consequent random addition~\cite{voss}, the correlated Levy
walks~\cite{shl}, etc.~\cite{czir}. We believe that from among the
above-mentioned methods, using the Markov chains is one of the
most important. We would like to demonstrate this statement in the
present paper.

In the next sections, the statistical properties of the
\textit{binary many-steps Markov chain} is examined. In spite of
the long-time history of studying the Markov sequences (see, for
example,~\cite{nag,kant,trib}), the concrete expressions for the
variance of sums of random variables in such strings have not been
obtained yet. Our model operates with two parameters governing the
conditional probability of the discrete Markov process,
specifically with the memory length $N$ and the correlation
parameter $\mu$. The correlation and distribution functions as
well as the variance $D$ being nonlinearly dependent on the length
$L$ of a word are derived analytically and calculated numerically.
The nonlinearity of the $D(L)$ function reflects the strong
correlations in the system. The evolved theory is applied to the
coarse-grained written texts and dictionaries, and to DNA strings
as well.

\section{Formulation of the problem}

\subsection{Markov Processes}

Let us consider a stationary binary sequence of symbols,
$a_{i}=\{0,1\}$. To determine the $N$-\textit{step Markov chain}
we have to introduce the conditional probability $P(a_{i}=0\mid
a_{i-N},a_{i-N+1},\dots ,a_{i-1})$ of following the definite
symbol (for example, zero) after symbols $a_{i-N},a_{i-N+1},\dots
,a_{i-1}$. Thus, it is necessary to define $2^{N}$ values of the
$P$-function corresponding to each possible configuration of the
symbols $a_{i-N},a_{i-N+1},\dots ,a_{i-1}$. We suppose that the
$P$-function has the form,
\begin{equation}
P(a_{i}=0\mid a_{i-N},a_{i-N+1},\dots ,a_{i-1})=\frac{1}{N}
\sum\limits_{k=1}^{N}f(a_{i-k},k).  \label{1}
\end{equation}
It is reasonable to assume the function $f$ to be decreasing with
an increase of the distance $k$ between the symbols $a_{i-k}$ and
$a_{i}$ in the Markov chain. However, for the sake of simplicity
we consider here a step-like memory function $f(a_{i-k},k)$
independent of the second argument $k$. As a result, our model is
characterized by three parameters only, specifically by $f(0)$,
$f(1)$, and $N$:
\begin{equation}
P(a_{i}=0\mid a_{i-N},a_{i-N+1},\dots ,a_{i-1})=\frac{1}{N}
\sum\limits_{k=1}^{N}f(a_{i-k}).  \label{2}
\end{equation}
Note that the probability $P$ in Eq.~(\ref{2}) depends on the
numbers of symbols 0 and 1 in the $N$-word but is independent of
the arrangement of the elements $a_{i-k}$. We also suppose that
\begin{equation}
f(0)+f(1)=1.  \label{2a}
\end{equation}
This relation provides the statistical equality of the numbers of
symbols zero and unity in the Markov chain under consideration. In
other words, the chain is non-biased. Indeed, taking into account
Eqs.~(\ref{2}) and (\ref{2a}) and the sequence of equations,
\begin{equation}
P(a_{i} = 1|a_{i-N},\dots ,a_{i-1})=1-P(a_{i}=0|a_{i-N},\dots
,a_{i-1})=\frac{1}{N}\sum\limits_{k=1}^{N}f(\tilde{a}_{i-N})=
P(a_{i}=0\mid \tilde{a} _{i-N},\dots ,\tilde{a}_{i-1}),
\end{equation}
one can see the symmetry with respect to interchange
$\tilde{a}_{i}\leftrightarrow a_{i}$ in the Markov chain. Here
$\tilde{a}_{i}$ is the symbol opposite to $a_{i}$,
$\tilde{a}_{i}=1-a_{i}$. Therefore, the probabilities of occurring
the words $(a_{1},\dots ,a_{L})$ and $(\tilde{a}_{1},\dots
,\tilde{a}_{L})$ are equal to each other for any word length $L$.
At $L=1$ this yields equal average probabilities of occurring $0$
and $1$ in the chain.

Taking into account the symmetry of a conditional probability $P$
with respect to a permutation of symbols $a_{i}$ (see
Eq.~(\ref{2})), we can simplify the notations and introduce the
conditional probability $p_{k}$ of occurring the symbol zero after
the $N$-word containing $k$ unities, e.g., after the word
$\underbrace{(11...1}_{k}\;\underbrace{00...0}_{N-k})$,
\begin{equation}
p_{k}=P(a_{N+1}=0\mid \underbrace{11\dots
1}_{k}\;\underbrace{00\dots 0} _{N-k})=\frac{1}{2}+\mu
(1-\frac{2k}{N}),  \label{14}
\end{equation}
with the correlation parameter $\mu $ being defined by the
relation
\begin{equation}
\mu =f(0)-\frac{1}{2}.  \label{3}
\end{equation}

We focus our attention on the region of $\mu $ determined by the
persistence inequality $0\leq \mu <1/2$. Nevertheless, the major
part of our results is valid for the anti-persistent region
$-1/2<\mu <0$ as well.

A similar rule for the production of an $N$-word $(a_{1},\dots
,a_{N})$ following after a word $(a_{0},a_1,\dots ,a_{N-1})$ was
suggested in Ref.~\cite{kant}. However, the conditional
probability $p_k$ of occurring the symbols $a_N$ does not depend
on the previous ones in the model~\cite{kant}.

\subsection{Statistical characteristics of the chain}

In order to investigate the statistical properties of the Markov
chain, we consider the distribution $W_{L}(k)$ of the words of
definite length $L$ by the number $k$ of unities in them,
\begin{equation}
k_{i}(L)=\sum\limits_{l=1}^{L}a_{i+l},  \label{5}
\end{equation}
and the variance of $k$,
\begin{equation}
D(L)=\overline{k^{2}}-\overline{k}^{2},  \label{7}
\end{equation}
where
\begin{equation}
\overline{f(k)}=\sum\limits_{k=0}^{L}f(k)W_{L}(k).  \label{8}
\end{equation}
If $\mu =0,$ one arrives at the known result for the
non-correlated Brownian diffusion,
\begin{equation}
D(L)=L/4.  \label{6}
\end{equation}
We will show that the distribution function $W_{L}(k)$ for the
sequence determined by Eq.~(\ref{14}) (with nonzero but not
extremely close to 1/2 parameter $\mu $) is the Gaussian with the
variance $D(L)$ nonlinearly dependent on $L$.

\subsection{Main equation}

For the stationary Markov chain, the probability
$b(a_{1}a_{2}\dots a_{N})$ of occurring a certain word
$(a_{1},a_{2},\dots ,a_{N})$ satisfies the equation,
\begin{equation}
b(a_{1}\dots a_{N})=\sum_{a=0,1}b(aa_{1}\dots a_{N-1})P(a_{N}\mid
a,a_{1},\dots ,a_{N-1}).  \label{10}
\end{equation}
Thus, we have $2^{N}$ homogeneous algebraic equations for $2^{N}$
words and the normalization equation $\sum b=1$. The set of
equations can be essentially simplified owing to the following
statement.

\textbf{Proposition} $\spadesuit$: \textit{The probability
$b(a_{1}a_{2}\dots a_{N})$ depends on the number $k$ of unities in
the $N$-word only but is independent of the arrangement of symbols
in the word $(a_{1},a_{2},\dots ,a_{N})$}.

This statement illustrated by Fig.~1 is valid owing to the chosen
simple model (\ref{2}), (\ref{14}) of the Markov chain. It can be
easily verified directly by the substitution of the obtained below
solution Eq.~(\ref{15}) into set (\ref{10}). Proposition
$\spadesuit$ leads to the very important property of isotropy: any
word $(a_{1},a_{2},\dots ,a_{L})$ appears with the same
probability as the inverted one, $(a_{L},a_{L-1},\dots ,a_{1})$.
\protect\begin{figure}[h!]
\begin{centering}
{\includegraphics{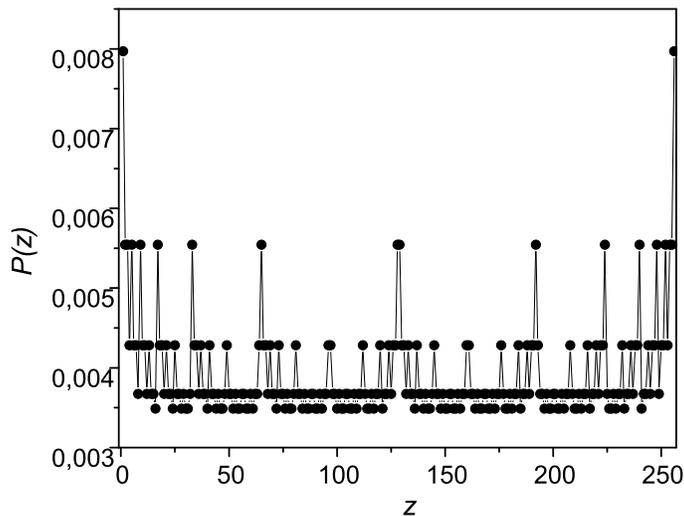}} \caption{The probability of occurring a
word $(a_1, a_2, \dots , a_ N)$ vs its number $z$ in the binary
code, $z=\sum_{i=1}^N a_i \cdot 2^{i-1}$ for $N=8$, $\mu=0.4$.}
\label{f1}
\end{centering}
\end{figure}

Let us apply set of Eqs.~(\ref{10}) to the word
$(\underbrace{11\dots 1}_{k}\;\underbrace{00\dots 0}_{N-k})$:
\begin{equation} \label{13}
b(\underbrace{11\dots 1}_{k}\;\underbrace{00\dots
0}_{N-k}) =b(0 \underbrace{11\dots 1}_{k}\;\underbrace{00\dots
0}_{N-k-1})p_{k}+b(1\underbrace{11\dots
1}_{k}\;\underbrace{00\dots 0}_{N-k-1})p_{k+1}.
\end{equation}
This yields the recursion relation for
$b(k)=b(\underbrace{11...1}_{k}\; \underbrace{00...0}_{N-k})$,
\begin{equation}
b(k)=\frac{1-p_{k-1}}{p_{k}}b(k-1)=\frac{N-2\mu (N-2k-2)}{N+2\mu
(N-2k)} b(k-1).  \label{15}
\end{equation}

\section{Distribution function of $L$-words}

In this section we investigate the statistical properties of the
Markov chain, specifically, the distribution of the words of
definite length $L$ by the number $k$ of unities in them. The
length $L$ can also be interpreted as the number of jumps of some
particle over an integer-valued 1-D lattice or as the time of the
diffusion imposed by the Markov chain under consideration. The
form of the distribution function $W_{L}(k)$ depends essentially
on the relation between the word length $L$ and the memory length
$N$. Therefore, we start our study with the simplest case, $L=N$.

\subsection{Statistics of $N$-words}

The value $b(k)$ is the probability that an $N$-word contains $k$
unities with a \textit{definite} order of symbols $a_i$.
Therefore, the probability $W_{N}(k)$ that an $N$-word contains
$k$ unities with \textit{arbitrary} order of symbols $a_i$ is
$b(k)$ multiplied by the number $\mathrm{C}_{N}^{k}=N!/k!(N-k)!$
of different permutations of $k$ unities in the $N$-word,
\begin{equation}
W_{N}(k)=\text{C}_{N}^{k}b(k).  \label{19}
\end{equation}
Combining Eqs.~(\ref{15}) and (\ref{19}), we get
\begin{equation}
W_{N}(k)= W_{N}(0)\text{C}_{N}^{k}\frac{\Gamma ( n+k) \Gamma (
n+N-k) }{\Gamma (n ) \Gamma (n+N)}  \label{18}
\end{equation}
with the parameter $n$ defined by
\begin{equation}\label{18a}
n= \frac{N(1-2\mu)}{4\mu}.
\end{equation}
The normalization constant $W_{N}(0)$ can be obtained from the
equality,
\begin{equation}
\sum\limits_{k=0}^{N}\text{C}_{N}^{k}b(k)=1.  \label{17}
\end{equation}
Note that the distribution $W_{N}(k)$ is an even function of the
variable $\kappa =k-N/2$,
\begin{equation} W_{N}(N-k)=W_{N}(k).  \label{19b}
\end{equation}
This fact is a direct consequence of the above-mentioned
statistical equivalency of zeros and unities in the Markov chain
under consideration. Let us analyze the distribution function
$W_{N}(k)$ for different relations between the parameters $N$ and
$\mu$.
\subsubsection{Limiting case of weak persistence, $n \gg 1$}

In the absence of correlations, $\mu \rightarrow 0$,
Eq.~(\ref{18}) and the Stirling formula yield the Gaussian
distribution at $k,\, N,\, N-k \gg 1$. In the most interesting
case of not too strong persistence, $n \gg 1$, one can also obtain
the Gaussian form for the distribution function,
\begin{equation}
W_{N}(k)=\frac{1}{\sqrt{2\pi D(N)}}\exp \left\{
-\frac{(k-N/2)^{2}}{2D(N)} \right\} ,  \label{27}
\end{equation}
with the $\mu$-dependent variance,
\begin{equation}
D(N)=\frac{N}{4(1-2\mu )}.  \label{28}
\end{equation}
Equation (\ref{27}) says that the $N$-words with equal numbers of
zeros and unities, $k=N/2$, are mostly probable. Note that the
persistence results in an increase of the variance $D(N)$ with
respect to its value $N/4$ at $\mu =0$. In other words, the
persistence is conductive to the intensification of the diffusion.
Inequality $n \gg 1$ provides $D(N) \ll N^{2}$. Therefore, despite
the increase of $D(N)$, the fluctuations of $(k-N/2)$ of the order
of $N$ are exponentially small.
\subsubsection{Intermediate case, $n \sim 1$}

If the parameter $n$ is an integer of the order of unity, the
distribution function $W_{N}(k)$ is a polynomial of degree
$2(n-1)$. In particular, at $n=1$, the function $W_{N}(k)$ is
constant,
\begin{equation}
W_{N}(k)=\frac{1}{N+1}.  \label{24}
\end{equation}
At $n\neq 1,$ $W_{N}(k)$ has a maximum at the middle of the
interval $[0,N]$.
\subsubsection{Limiting case of strong persistence, $n \ll 1$}

In this situation, $W_{N}(k)$ assumes the maximal values at $k=0$
and $k=N$,
\begin{equation}
W_{N}(1)=W_{N}(0)\frac{nN}{N-1} \ll W_{N}(0). \label{20}
\end{equation}
Formula (\ref{20}) describes the sharply decreasing $W_{N}(k)$ as
$k$ changes from $0$ to $1$ (and from $N$ to $N-1$). Then, at
$1<k<N/2$, the function $W_{N}(k)$ decreases more slowly with an
increase in $k$,
\begin{equation}
W_{N}(k)=W_{N}(0)\frac{nN}{k(N-k)}.  \label{21}
\end{equation}
At $k=N/2,$ the probability $W_{N}(k)$ achieves its minimal value,
\begin{equation}
W_{N}\left(\frac{N}{2}\right)=W_{N}(0)\frac{4n}{N}. \label{22}
\end{equation}

The values $W_{N}(0)=W_N (N)$ are nearly $1/2$ to a logarithmic
accuracy.

The evolution of the distribution function $W_N(k)$ from the
Gaussian form to the inverse one with a decrease of the parameter
$n$ is shown in Fig.~2. Below we restrict ourselves to the case of
weak persistence, $n \gg 1$.

Formulas (\ref{27}) and (\ref{28}) describe the statistical
properties of $L$-words for the fixed ''diffusion time'' $L=N$. It
is necessary to look into the distribution function $W_{L}(k)$ for
the general situation, $L\neq N$. We start the analysis with
$L<N$.
\protect\begin{figure}[h!]
\begin{centering}
{\includegraphics{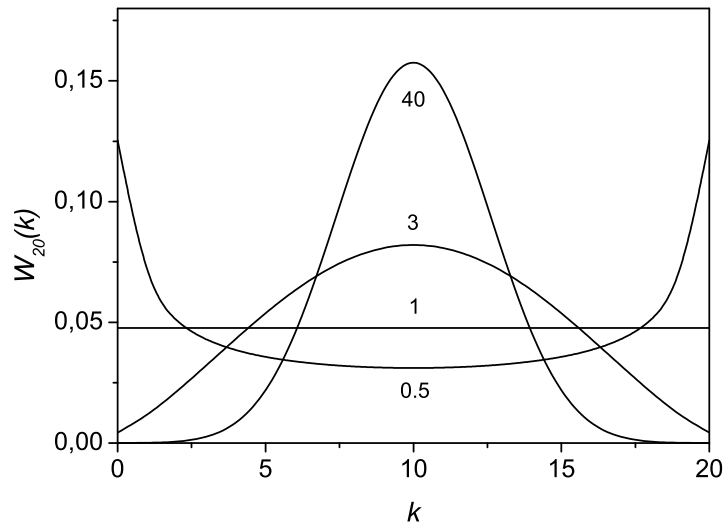}} \caption{The distribution function
$W_N(k)$ for $N$=20 and different values of the parameter $n$
shown near the curves.} \label{f2}
\end{centering}
\end{figure}

\subsection{Statistics of $L$-words with $L<N$}
\subsubsection{Distribution function $W_{L}(k)$}

The distribution function $W_{L}(k)$ at $L<N$ can be given as
\begin{equation}
W_{L}(k)=\sum\limits_{i=k}^{k+N-L}b(i)\text{C}_{L}^{k}\text{C}_{N-L}^{i-k}.
\label{29}
\end{equation}
This equation follows from the consideration of $N$-words
consisting of two parts,
\begin{equation}
(\underbrace{a_{1},\dots ,a_{N-L},}_{i-k\text{
unities}}\;\underbrace{a_{N-L+1},\dots ,a_{N}}_{k \text{
unities}}). \label{29b}
\end{equation}
The total number of unities in this word is $i$. The right-hand
part of the word ($L$-sub-word) contains $k$ unities. The
remaining $i-k$ unities are situated within the left-hand part of
the word ($(N-L)$-sub-word). The multiplier
$\mathrm{C}_{L}^{k}\mathrm{C}_{N-L}^{i-k}$ in Eq.~(\ref{29}) takes
into account all possible permutations of the symbols ''1'' within
the $N$-word on condition that the $L$-sub-word always contains
$k$ unities. Then we perform the summation over all possible
values of the number $i$. Note that Eq.~(\ref{29}) is valid due to
the main proposition $\spadesuit$ formulated in Subsec.~C of the
previous section.

In this subsection, we focus our attention on the most important
limiting case $n,\;k,\;L-k \gg 1$. The straightforward
calculations with using the Stirling formula give the result,
\begin{equation}
W_{L}(k)=\frac{1}{\sqrt{2\pi D(L)}}\exp \left\{
-\frac{(k-L/2)^{2}}{2D(L)} \right\}  \label{31}
\end{equation}
with
\begin{equation}
D(L)=\frac{L}{4}(1+mL), \quad m= \frac{1}{2n}=\frac{2\mu}{N(1-2\mu
)}. \label{32}
\end{equation}

The last equation allows one to analyze the behavior of the
variance $D(L)$ with an increase in the ``diffusion time'' $L$. At
small $mL \ll 1$ the dependence $D(L)$ follows the classical law
of the Brownian diffusion, $D(L)\approx L/4$. Then, at $mL\sim 1$,
the function $D(L)$ becomes super-linear and meets the value
(\ref{28}) at $L=N$.

Such an unusual behavior of the variance $D(L)$ raises an issue as
to what particular type of the diffusion equation corresponds to
the nonlinear dependence $D(L)$ in Eq.~(\ref{32}). In the
following subsection, when solving this problem, we will obtain
the conditional probability $p^{(0)}$ of occurring the symbol zero
after a given $L$-word with $L<N$. The ability to find $p^{(0)}$,
with some reduced information about preceding symbols being
available, is very important for the study of the self-similarity
of the Markov chain (see Subsubsec.~3 of this Subsection).
\subsubsection{Generalized diffusion equation}

It is quite obvious that the distribution $W_{L}(k)$ satisfies the
equation
\begin{equation}
W_{L+1}(k)=W_{L}(k)p^{(0)}(k)+W_{L}(k-1)p^{(1)}(k-1).  \label{33}
\end{equation}
Here $p^{(0)}$ is the probability of occurring ''0'' after an
average-statistical $L$-word containing $k$ unities and $p^{(1)}$
is the probability of occurring ''1'' after an $L$-word containing
$k-1$ unities. At $L<N$, the probability $p^{(0)}$ can be written
as
\begin{equation}
p^{(0)}=\frac{1}{W_L(k)}
\sum\limits_{i=k}^{k+N-L}p_{i}b(i)\mathrm{C}_{L}^{k}\mathrm{C}_{N-L}^{i-k}.
\label{34}
\end{equation}
The product $b(i)\mathrm{C}_{L}^{k}\mathrm{C}_{N-L}^{i-k}$ in this
formula represents the conditional probability of occurring the
$N$-word containing $i$ unities, the right-hand part of which, the
$L$-sub-word, contains $k$ unities (compare with Eqs.~(\ref{29}),
(\ref{29b})).

The product $b(i)\mathrm{C}_{N-L}^{i-k}$ in Eq.~(\ref{34}) is a
sharp function of $i$ with a maximum at some point $i=i_0$ whereas
$p_{i}$ obeys linear law (\ref{14}). This implies that $p_{i}$ can
be factored out of the summation sign being taken at point
$i=i_0$. The asymptotical calculation shows that point $i_0$ is
described by the equation,
\begin{equation}
i_{0}=\frac{N}{2}-\frac{L/2}{1-2\mu (1-L/N)}\left(
1-\frac{2k}{L}\right). \label{35}
\end{equation}
Expression (\ref{14}) taken at point $k=i_0$ gives the desired
formula for $p^{(0)}$ because
\begin{equation}
\sum\limits_{i=k}^{k+N-L}b(i)\mathrm{C}_{L}^{k}\mathrm{C}_{N-L}^{i-k}
\end{equation}
is obviously equal to $W_L(k)$. Thus, we have
\begin{equation}
p^{(0)}(k)=\frac{1}{2}+\frac{\mu L}{N-2\mu (N-L)}\left(
1-\frac{2k}{L}\right). \label{36}
\end{equation}

Let us consider a very important fact following from
Eq.~(\ref{35}). If the concentration of unities in the right-hand
part of the word (\ref{29b}) is higher than $1/2$, $k/L >1/2$,
then the most probable concentration $(i_0-k)/(N-L)$ of unities in
the left-hand part of this word is likewise increased,
$(i_0-k)/(N-L)>1/2$. At the same time, the concentration
$(i_0-k)/(N-L)$ is less than $k/L$,
\begin{equation}\label{36b}
\frac{1}{2} <\frac{i_0-k}{N-L}<\frac{k}{L}.
\end{equation}
This means that the increased concentration of unities in the
$L$-words is necessarily accompanied by the existence of a certain
tail with an increased concentration of unities as well. We name
such a phenomenon as the \textit{macro-persistence}. An analysis
performed in the next section will indicate that the correlation
length $l_c$ of this tail is $\gamma N $ with $\gamma \geq 1$
dependent on the parameter $\mu$ only. It is evident from the
above-mentioned property of the isotropy of the Markov chain that
there are two correlation tails from the both sides of the
$L$-word.

By going over to the continuous limit in Eq.~(\ref{33}) and using
Eq.~(\ref{36}) and the relation $p^{(1)}(k-1)=1-p^{(0)}(k-1)$, we
obtain the diffusion equation generalized to the case of the
correlated Markov process,
\begin{equation}
\frac{\partial W}{\partial L}=\frac{1}{8}\frac{\partial
^{2}W}{\partial \kappa ^{2}}-\eta(L)\left( \kappa \frac{\partial
W}{\partial \kappa }+W\right),  \label{39}
\end{equation}
with $\kappa =k-L/2$ and
\begin{equation}\label{39b}
\eta(L)=\frac{2\mu}{(1-2\mu )N+2\mu L}.
\end{equation}
Equation (\ref{39}) has a solution of the Gaussian form
Eq.~(\ref{31}) with the variance $D(L)$ satisfying the ordinary
differential equation,
\begin{equation}
\frac{\mathrm{d}D}{\mathrm{d}L}=\frac{1}{4}+2\eta(L)D. \label{40}
\end{equation}
Its solution with the boundary condition $D(0)=0$ coincides with
(\ref {32}).
\subsubsection{Self-similarity of the persistent Brownian diffusion}

In this subsection we point out one of the most important
properties of the Markov chain being considered, namely, its
self-similarity. Let us reduce the $N$-step Markov sequence by
regularly (or randomly) removing some symbols and introduce the
decimation parameter $\lambda$,
\begin{equation}
\lambda =N^{\ast }/N \leq 1.  \label{41}
\end{equation}
Here $N^{\ast }$ is a renormalized memory length for the reduced
$N^{\ast }$-step Markov chain. According to Eq.~(\ref{36}), the
conditional probability $p_{k}^{\ast }$ of occurring the symbol
zero after $k$ unities among the preceding $N^{\ast }$ symbols is
described by the formula,
\begin{equation}
p_{k}^{\ast }=\frac{1}{2}+\mu ^{\ast }\left( 1-\frac{2k}{N^{\ast
}}\right), \label{42}
\end{equation}
with
\begin{equation}
\mu ^{\ast }=\mu \frac{\lambda }{1-2\mu (1-\lambda )}.  \label{43}
\end{equation}
The comparison of Eqs.~(\ref{14}) and (\ref{42}) shows that the
reduced chain possesses the same statistical properties as the
initial one but is characterized by the renormalized parameters
($N^{\ast }$, $\mu ^{\ast }$) instead of ($N$, $\mu $). Thus,
Eqs.~(\ref{41}) and (\ref{43}) determine the one-parametrical
renormalization of the parameters of the stochastic process
defined by Eq.~(\ref{14}).

The astonishing property of the reduced sequence consists in that
\textit{the variance $D^{\ast }(L)$ is invariant with respect to
the one-parametric decimation transformation (\ref{41}),
(\ref{43})}. Therefore, it coincides with the function $D(L)$ for
the initial Markov chain:
\begin{equation} \label{44}
D^{\ast }(L) = \frac{L}{4}\left( 1+\frac{2\mu ^{\ast }}{1-2\mu
^{\ast }} \frac{L}{N^{\ast }}\right) = \frac{L}{4}\left(
1+\frac{2\mu }{1-2\mu }\frac{L}{N}\right) =D(L).
\end{equation}
The invariance of the function $D(L)$ at $L<N$ was referred by us
to as the phenomenon of self-similarity. It is demonstrated in
Fig.~3 and is also discussed in Sec.~IV A.
\protect\begin{figure}[h!]
\begin{centering}
{\includegraphics{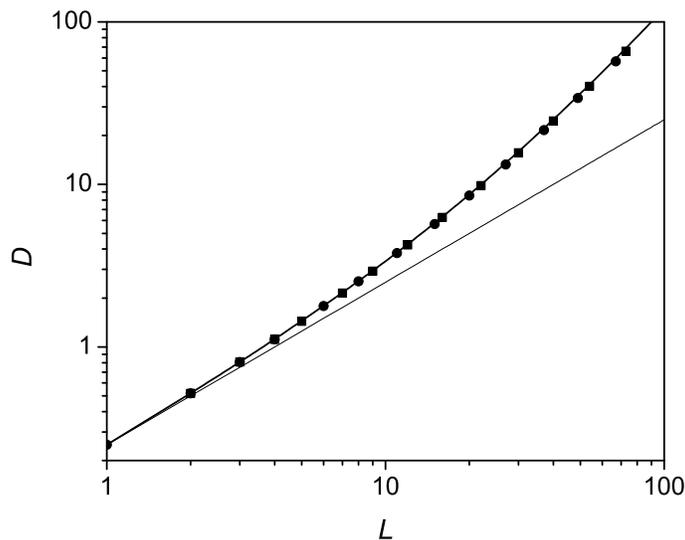}} \caption{The dependence of the variance
$D$ on the tuple length $L$ for the generated sequence with
$N=100$ and $\mu=0.4$ (solid line) and for the decimated sequences
(the parameter of decimation $\lambda =0.5$). Squares and circles
correspond to the stochastic and deterministic reduction,
respectively. The thin solid line describes the non-correlated
Brownian diffusion, $D(L)=L/4$.} \label{f3}
\end{centering}
\end{figure}

\subsection{Long-range diffusion, $L>N$}

Unfortunately, the very useful proposition $\spadesuit$ is valid
for the words of the length $L\leq N$ only and cannot be applied
to the analysis of the long words with $L>N$. Therefore,
investigating the statistical properties of the long words
represents a rather challenging combinatorial problem and requires
new physical approaches for its simplification. Thus, we start
this subsection by analyzing the correlation properties of the
long words.
\subsubsection{Correlation length}

Let us rewrite Eq.~(\ref{14}) in the form,
\begin{equation}
<a_{i+1}>=\frac{1}{2}+\mu \left(
\frac{2}{N}\sum_{k=i-N+1}^{i}<a_{k}>-1 \right).  \label{46}
\end{equation}
The angle brackets denote the averaging of the density of unities
in some region of the Markov chain for its definite realization.
The averaging is performed over distances much greater than unity
but far less than the memory length $N$. Note that this averaging
differs from the statistical averaging over the ensemble of
realizations of the Markov chain denoted by the bar in
Eqs.~(\ref{7}) and (\ref{8}). Equation (\ref{46}) is a
relationship between the average densities of unities in two
different macroscopic regions of the Markov chain, namely, in the
vicinity of $(i+1)$-th element and in the region $(i-N,\,\,i)$.
Such an approach is similar to the mean field approximation in the
theory of the phase transitions and is asymptotically exact under
the condition $N\rightarrow \infty$. In the continuous limit,
Eq.~(\ref{46}) can be rewritten in the integral form,
\begin{equation}
<a(i)>=\frac{1}{2}+\mu \left(
\frac{2}{N}\int_{i-N}^{i}<a(k)>\textrm{d}k-1\right). \label{47}
\end{equation}
It has the obvious solution,
\begin{equation}
<a(i)-\frac{1}{2}>=<a(0)-\frac{1}{2}>\exp \left( i/\gamma
N\right),  \label{49}
\end{equation}
where $\gamma $ is determined by the relation,
\begin{equation}
\gamma \left( \exp \left( \frac{1}{\gamma }\right) -1\right)
=\frac{1}{2\mu}.  \label{50}
\end{equation}
The last equation has a unique solution $\gamma (\mu )$ for any
value of $\mu \in(0, 1/2)$.

Formula (\ref{49}) shows that any fluctuation (the difference
between $<a(i)>$ and the equilibrium value of
$\overline{a_i}=1/2$) is exponentially damped at distances of the
order of the \textit{correlation length} $l_c$,
\begin{equation}\label{50b}
l_{c}=\gamma N.
\end{equation}
Law (\ref{49}) describes the phenomenon of the \textit{persistent
macroscopic correlations} discussed in the previous subsection.
This phenomenon is governed by both of the parameters, the memory
length $N$ and the persistence parameter $\mu$. According to
Eqs.~(\ref{50}), (\ref{50b}), the correlation length $l_c$ goes
logarithmically to infinity with an increase in $\mu$, at $\mu
\rightarrow 1/2$. At $\mu \rightarrow 0$, the macro-persistence is
broken and the correlation length tends to zero.
\subsubsection{Correlation function}

Using the already studied correlation properties of the the Markov
sequence and some heuristic reasons, one can obtain the
correlation function ${\cal K}(r)$,
\begin{equation}
{\cal K}(r)=\overline{a_{i}a_{i+r}}-\overline{a_{i}}^2, \label{51}
\end{equation}
and then the variance $D(L)$,
\begin{equation}
D(L)=\overline{\left( \sum a_{i}\right) ^{2}}-\left(
\overline{\sum a_{i}} \right) ^{2}.  \label{51b}
\end{equation}
Comparing Eqs.~(\ref{51}) and (\ref{51b}) and taking into account
the unbiased property of the sequence, $\overline{a_{i}}=1/2$, it
is easy to derive the general relationship between the functions
${\cal K}(r)$ and $D(L)$,
\begin{equation} \label{51c}
D(L)=\frac{L^{2}}{4}+4\sum_{i=1}^{L-1}\sum_{r=1}^{L-i}{\cal K}(r).
\end{equation}
Considering (\ref{51c}) as an equation with respect to ${\cal
K}(r)$, one can find its solution given as
\[
{\cal K}(1) = \frac{1}{2}D(2)-\frac{1}{4}, \quad {\cal K}(2) =
\frac{1}{2}D(3)-D(2)+\frac{1}{8},
\]
\begin{equation}\label{51e}
{\cal K}(r)=\frac{1}{2}\left[D(r+1) -2D(r) +D(r-1)\right], \quad
r\geq 3.
\end{equation}
This solution has a very simple form in the continuous limit,
\begin{equation}\label{51f}
{\cal K}(r) = \frac{1}{2}\frac{{\textrm d}^2 D(r)}{{\textrm
d}r^2}.
\end{equation}

Equations~(\ref{51e}) and (\ref{32}) give the correlation function
at $r<N$,
\begin{equation}
{\cal K}(r)=C_r \frac{2\mu }{1-2\mu }\frac{1}{N} =C_{r}m, \qquad
C_{1}=1/2, \quad C_{2}=1/8, \quad C_{3\leq r\leq N}=1/4,
\label{53}
\end{equation}
or
\begin{equation}
{\cal K}(r)=\frac{m}{4 }, \qquad r \leq N \label{54b}
\end{equation}
in the continuous approximation. The independence of the
correlation function of $r$ at $r<N$ results from our choice of
the conditional probability in the simplest form (\ref{14}). At
$r>N$, the function ${\cal K}(r)$ should decrease because of loss
of the memory. Therefore, based on Eqs.~(\ref{49}) and
(\ref{50b}), let us prolongate the correlator ${\cal K}(r)$ as the
exponentially decreasing function at $r>N$,
\begin{equation}
{\cal K}(r)=\frac{m}{4}\cases {1,\;\qquad \;\;\;\;\;\qquad r\leq
N, \cr \exp \left(-\frac{r-N}{l_{c}}\right ), \;\;r>N.} \label{55}
\end{equation}
Correspondingly, the variance $D(L)$ becomes
\begin{equation} \label{56}
D(L)=\frac{L}{4}\left(1+m F(L)\right),
\end{equation}
with
\begin{equation} \label{57}
F(L)= \cases {L, \qquad \qquad \qquad \qquad  \quad \qquad \qquad
\qquad \qquad \qquad \qquad \qquad \qquad L<N, \cr 2(1+ \gamma)N -
(1+2\gamma ) \frac{N^2}{L}- 2\gamma^{2}\frac{N^2}{L} \left[1-\exp
\left( -\frac{L-N}{l_c}\right) \right], \, \qquad L>N.}
\end{equation}

The plot of Eq.~(\ref{56}) for $N=100$ and $\mu=0.4$ is shown by
the solid line in Fig.~4. For comparison, the straight line in the
figure corresponds to the dependence $D(L)=L/4$ for the usual
Brownian diffusion without correlations (for $\mu=0$). It is
clearly seen that the plot of variance (\ref{56}) contains two
qualitatively different portions. One of them, at $L\sim N$, is
the super-linear curve that moves away from the line $D=L/4$ with
an increase of $L$ as a result of the persistence. For $L\gg N$,
the plot $D(L)$ achieves the linear asymptotics,
\begin{equation}\label{58}
D(L)\cong \frac{L}{4}(1+2(1+\gamma)mN).
\end{equation}
This phenomenon can be explained as a result of the diffusion
where each practically \textit{independent} step $\sim
D^{1/2}(N+l_c)$ of wandering represents a path traversed during
the characteristic ``fluctuating time'' $\Delta L \sim (N+l_c)$.
Since these steps of wandering are quasi-independent, the
distribution function $W_L(k)$ is the Gaussian not only at $L<N$
(see Eq.~(\ref{31})) but also in the case $L>N, \, l_c$.
\protect\begin{figure}[h!]
\begin{centering}
{\includegraphics{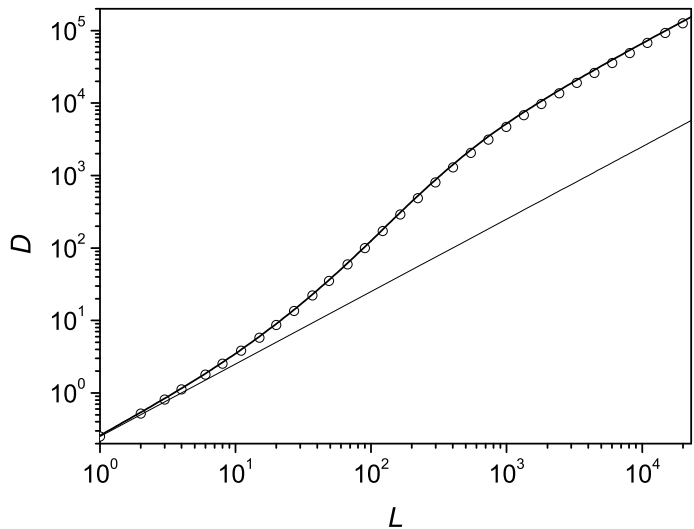}} \caption{The numerical simulation of
the dependence $D(L)$ for the generated sequence with $N=100$ and
$\mu=0.4$ (circles). The solid line is the plot of function
Eq.~(55) with the same values of $N$ and $\mu$.} \label{f4}
\end{centering}
\end{figure}

Note that the above-mentioned phenomenon of the self-similarity
relates only to the portion $L<N$ of the curve $D(L)$. Since the
decimation procedure leads to the decrease of the parameter $\mu$
(see Eq.~(\ref{43})), the plot of asymptotics (\ref{58}) for the
reduced sequence at $L\gg N^{\ast}$ goes below the plot for the
initial chain.

\section{Results of
numerical simulations and applications}

In this section, we support the obtained above analytical results
by numerical simulations of the Markov chain with the conditional
probability Eq.~(\ref{14}). Besides, the properties of the studied
binary $N$-step Markov chain are compared with ones for the
natural objects, specifically for the coarse-grained written and
DNA texts.

\subsection{Numerical simulations of the Markov chain}

The first stage of the construction of the $N$-step Markov chain
was a generation of the initial non-correlated $N$ symbols, zeros
and unities, identically distributed with equal probabilities 1/2.
Each consequent symbol was then added to the chain with the
conditional probability determined by the previous $N$ symbols in
accordance with Eq.~(\ref{14}). Than we numerically calculated the
variance $D(L)$ by means of Eq.~(\ref{7}). The circles in Fig.~4
represent the calculated variance $D(L)$ for the 100-step Markov
chain generated at $\mu =0.4$. A very good agreement between the
analytical result in Eq.~(\ref{56}) and the numerical simulation
can be observed.

The numerical simulation was also used for the demonstration of
the proposition $\spadesuit$ (Fig.~1) and the self-similarity
property of the Markov sequence (Fig.~3). The squares in Fig.~3
represent the variance $D(L)$ for the sequence obtained by the
stochastic decimation of the initial Markov chain (solid line)
where each symbol was omitted with the probability 1/2. The
circles in this figure correspond to the regular reduction of the
sequence by removing each second symbol.

And finally, the numerical simulations have allowed us to make
sure that we are able to determine the parameters $N$ and $\mu$ of
a given binary sequence. We generated Markov sequences with
different parameters $N$ and $\mu$ and defined numerically the
corresponding curves $D(L)$. Then we solved the inverse problem of
the reconstruction of the parameters $N$ and $\mu$ by analyzing
the curves $D(L)$. The reconstructed parameters were always in a
good agreement with their prescribed values. In the next
subsections we apply this ability to the treatment of the
statistical properties of literary and DNA texts.

\subsection{Literary texts}

It is well-known that the statistical properties of the
coarse-grained texts written in any language show a remarkable
deviation from random sequences~\cite{schen,kant}. In order to
check the applicability of the theory of the binary $N$-step
Markov chains to literary texts we resorted to the procedure of
coarse graining by the random mapping of all characters of the
text onto binary set of symbols, zeros and unities. The
statistical properties of the coarse-grained texts depend, but not
significantly, on the kind of mapping. This is illustrated by the
curves in Fig.~5 where the variance $D(L)$ for five different
kinds of the mapping of Bible is presented. Usually, the random
mapping leads to nonequal numbers of unities and zeros, $k_1$ and
$k_0$, in the coarse-grained sequence. The particular analysis
shows that the variance $D(L)$ (\ref{32}) gets the additional
multiplier,
\[
\frac{4k_0 k_1}{(k_0+k_1)^2},
\]
in this biased case. In order to derive the function $D(L)$ for
the non-biased sequence, we divided the numerically calculated
value of the variance by this multiplier.
\protect\begin{figure}[h!]
\begin{centering}
{\includegraphics{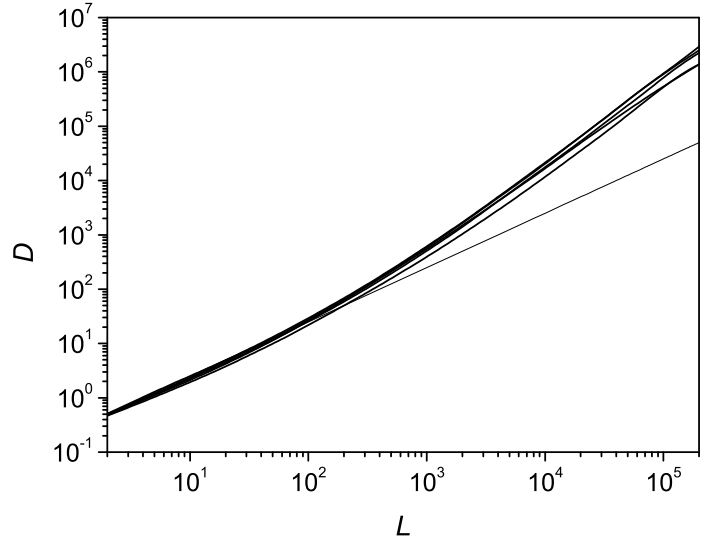}} \caption{The dependence $D(L)$ for the
coarse-grained text of the Bible obtained by means of five
different kinds of random mapping.} \label{f5}
\end{centering}
\end{figure}

The study of different written texts has shown that all of them
are featured by the pronounced persistent correlations. It is
demonstrated by Fig.~6 where five variance curves go significantly
higher than the straight line $D=L/4$. However, it should be
emphasized that regardless of the kind of mapping the initial
portions, $L<80$, of the curves correspond to a slight
anti-persistent behavior (see insert to Fig.~7). Moreover, for
some inappropriate kinds of mapping (e.g., when all vowels are
mapped onto the same symbol) the anti-persistent portions can
reach the values of $L\sim 1000$. In order to avoid this problem,
all the curves in Fig.~6 are obtained for the definite
representative mapping: (a-m) $\rightarrow$ 0; (n-z) $\rightarrow$
1. \protect\begin{figure}[h!]
\begin{centering}
{\includegraphics{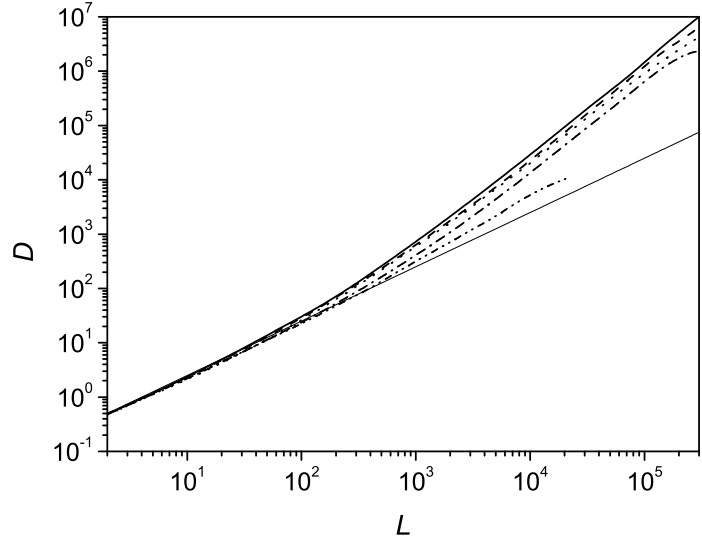}} \caption{The dependence $D(L)$ for the
coarse-grained texts of collection of works on the computer
science ($m=2.2\cdot 10^{-3}$, solid line), Bible in Russian
($m=1.9\cdot 10^{-3}$, dashed line), Bible in English ($m=1.5\cdot
10^{-3}$, dotted line), ``History of Russians in the 20-th
Century'' by Oleg Platonov ($m=6.4\cdot 10^{-4}$, dash-dotted
line), and ``Alice's Adventures in Wonderland'' by Lewis Carroll
($m=2.7\cdot 10^{-4}$, dash-dot-dotted line).} \label{f6}
\end{centering}
\end{figure}

Thus, the persistence is the common property of the binary
$N$-step Markov chains and the coarse-grained written texts at
large scales. Moreover, the written texts as well as the Markov
sequences possess the property of the self-similarity. Indeed, the
curves in Fig.~7 obtained from the text of Bible with different
levels of the deterministic decimation demonstrate the
self-similarity. Presumably, this property is the mathematical
reflection of the well-known hierarchy in the linguistics:
\textit{letters $\rightarrow$ syllables $\rightarrow$ words
$\rightarrow$ sentences $\rightarrow$ paragraphs $\rightarrow$
chapters $\rightarrow$ books $\rightarrow $ collected works}.
\protect\begin{figure}[h!]
\begin{centering}
{\includegraphics{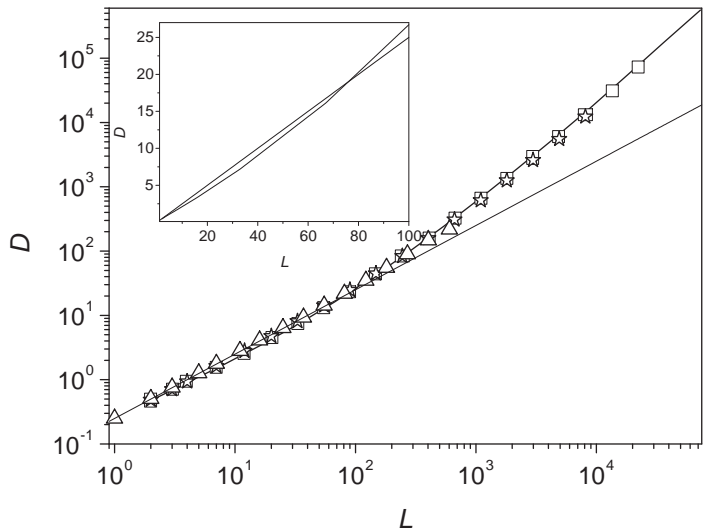}} \caption{The dependence of the variance
$D$ on the tuple length $L$ for the coarse-grained text of Bible
(solid line) and for the decimated sequences with different
parameters $\lambda$: $\lambda = 3/4 $ (squares), $\lambda = 1/2 $
(stars), and $\lambda = 1/256 $ (triangles). The insert
demonstrate the anti-persistent portion of the $D(L)$ plot for
Bible.}\label{f7}
\end{centering}
\end{figure}

All the above-mentioned circumstances allow us to suppose that our
theory of the binary $N$-step Markov chains can be applied to the
description of the statistical properties of the texts of natural
languages. However, in contrast to the generated Markov sequence
(see Fig.~4) where the full length $\mathcal{M}$ of the chain is
far greater than the memory length $N$, the coarse-grained texts
described by Fig.~6 are of relatively short length
$\mathcal{M}<N$. In other words, the coarse-grained texts are
similar not to the Markov chains but rather to some non-stationary
short fragments. This implies that each of the written texts is
correlated throughout the whole of its length. Therefore, for the
written texts, it is impossible to observe the second portion of
the curve $D(L)$ parallel (in the log-log scale) to the line
$D(L)=L/4$, similar to that shown in Fig.~4. As a result, one
cannot define the values of the both parameters $N$ and $\mu$ for
the coarse-grained texts. The analysis of the curves in Fig.~6 can
give the combination $m=2\mu/N(1-2\mu)$ only (see Eq.~(\ref{32})).
Perhaps, this particular combination is the real parameter
governing the persistent properties of the literary texts.

We would like to note that the origin of the long-range
correlations in the literary texts is hardly related to the
grammatical rules as is claimed in Ref.~\cite{kant}. At short
scales $L\leq 80$ where the grammatical rules are in fact
applicable the character of correlations is anti-persistent
whereas semantic correlations lead to the global persistent
behavior of the variance $D(L)$ throughout the whole of literary
text.

The numerical estimations of the persistent parameter $m$ and the
characterization of the languages and different authors using this
parameter can be regarded as a new intriguing problem of the
linguistics. For instance, the unprecedented low value of $m$ for
the very inventive work by Lewis Carroll as well as the closeness
of $m$ for the texts of English and Russian versions of Bible are
of certain interest.

It should be noted that there exist special kinds of short-range
correlated texts which can be specified by both of the parameters,
$N$ and $\mu$. For example, all dictionaries consist of the
families of words where some preferable letters are repeated more
frequently than in their other parts. Yet another example of the
shortly correlated texts is any lexicographically ordered list of
words. The analysis of written texts of this kind is given below.

\subsection{Dictionaries}

As an example, we have investigated the statistical properties of
the coarse-grained alphabetical list of the most frequently used
15462 English words. In contrast to other texts, the statistical
properties of the coarse-grained dictionaries are very sensitive
to the kind of mapping. If one uses the above-mentioned mapping,
(a-m) $\rightarrow$ 0; (n-z) $\rightarrow$ 1, the behavior of the
variance $D(L)$ similar to that shown in Fig.~6 would be obtained.
The particular construction of the dictionary manifests itself if
the preferable letters in the neighboring families of words are
mapped onto the different symbols. The variance $D(L)$ for the
dictionary coarse-grained by means of such mapping is shown by
circles in Fig.~8. It is clearly seen that the graph of the
function $D(L)$ consists of two portions similarly to the curve in
Fig.~4 obtained for the generated $N$-step Markov sequence. The
fitting of the curve in Fig.~8 by the function (\ref{56}) (solid
line in Fig.~8) yielded the values of the parameters $N=180$ and
$\mu =0.44$. The parameter $\gamma$ given by Eq.~(\ref{50}) is
around 3.9. Note that the characteristic fluctuation length
$N(1+\gamma)$ for these $N$ and $\gamma$ is nearly 900. This value
corresponds qualitatively to the length of the family of words in
the dictionary. \protect\begin{figure}[h!]
\begin{centering}
{\includegraphics{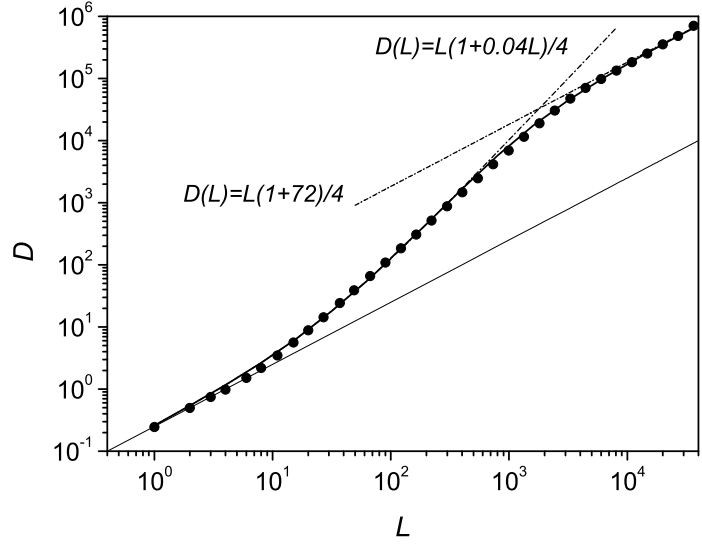}} \caption{The dependence $D(L)$ for the
coarse-grained alphabetical list of 15462 English words (circles).
The solid line is the plot of function Eq.~(55) with the fitting
parameters $N=180$ and $\mu=0.44$.} \label{f8}
\end{centering}
\end{figure}

\subsection{DNA texts}

It is known that any DNA text is written by four ``characters'',
specifically by adenine (A), cytosine (C), guanine (G), and
thymine (T). Therefore, there are three nonequivalent types of the
DNA text mapping onto one-dimensional binary sequences of zeros
and unities. The first of them is the so-called purine-pyrimidine
rule, \{A,G\} $\rightarrow$ 0, \{C,T\} $\rightarrow$ 1. The second
one is the hydrogen-bond rule, \{A,T\} $\rightarrow$ 0, \{C,G\}
$\rightarrow$ 1. And, finally, the third is \{A,C\} $\rightarrow$
0, \{G,T\} $\rightarrow$ 1. \protect\begin{figure}[h!]
\begin{centering}
{\includegraphics{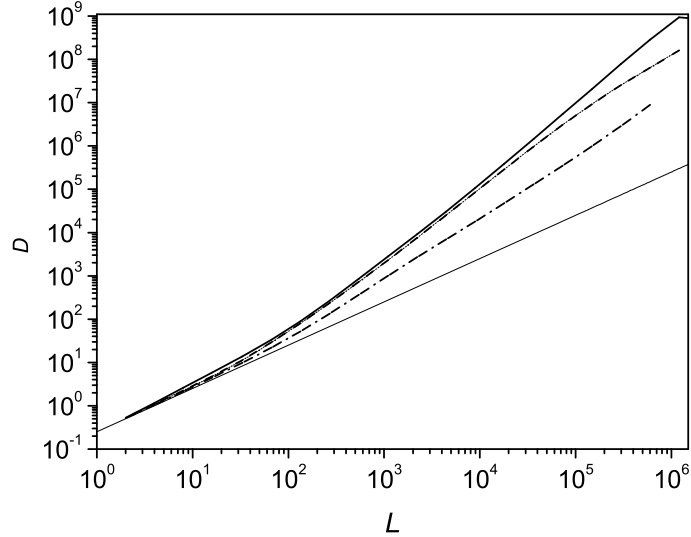}} \caption{The dependence $D(L)$ for the
coarse-grained DNA text of \textit{Bacillus subtilis, complete
genome}, for three nonequivalent kinds of the mapping. Solid,
dashed, and dash-dotted lines correspond to the mappings \{A,G\}
$\rightarrow$ 0, \{C,T\} $\rightarrow$ 1 (the parameter
$m=4.1\cdot 10^{-2}$), \{A,T\} $\rightarrow$ 0, \{C,G\}
$\rightarrow$ 1 ($m=2.5\cdot 10^{-2}$), and \{A,C\} $\rightarrow$
0, \{G,T\} $\rightarrow$ 1 ($m=1.5\cdot 10^{-2}$), respectively.}
\label{f9}
\end{centering}
\end{figure}

By way of example, the variance $D(L)$ for the coarse-grained text
of \textit{Bacillus subtilis, complete genome}
(ftp:$//$ftp.ncbi.nih.gov$/$genomes$/$bacteria$/$bacillus\_subtilis$/$NC\_000964.gbk)
is displayed in Fig.~9 for all possible types of mapping. One can
see that the persistent properties of DNA are more pronounced than
for the written texts and, contrary to the written texts, the
$D(L)$ dependence for DNA does not exhibit the anti-persistent
behavior at small values of $L$. However, as well as for the
written texts, the $D(L)$ curve for DNA does not contain the
linear portion given by Eq.~(\ref{58}). This suggests that the DNA
chain is not a stationary sequence. In this connection, we would
like to point out that the DNA texts represent the collection of
extended non-coding regions interrupted by small coding regions
(see, for example,~\cite{prov}). According to Fig.~9, the coding
regions do not interrupt the correlation between the non-coding
areas, and the DNA system is fully correlated throughout its whole
length.

The noticeable deviation of different curves in Fig.~9 from each
other demonstrate, in our opinion, that the DNA texts are much
more complex objects in comparison with the written ones. Indeed,
the different kinds of mapping reveal and emphasize various types
of physical attractive correlations between the nucleotides in
DNA, such as the strong purine-purine and pyrimidine-pyrimidine
persistent correlations (the upper curve), and the correlations
caused by a weaker attraction A$\leftrightarrow$T and
C$\leftrightarrow$G (the middle curve).

\section{Conclusion}

Thus, we have developed a new approach for the description of the
strongly correlated one-dimensional systems. The simple exactly
solvable model of the uniform binary $N$-step Markov chain is
presented. The memory length $N$ and the parameter $\mu$ of the
persistent correlations are two parameters in our theory. Usually,
the correlation function ${\cal K}(r)$ is employed as the input
characteristics for the description of the correlated random
systems. Yet, the function ${\cal K}(r)$ describes not only the
direct interconnection of the elements $a_i$ and $a_{i+r}$, but
also takes into account their indirect interaction via other
elements. Since our approach operates with the ``origin''
parameters $N$ and $\mu$, we believe that it allows us to disclose
the intrinsic properties of the system which provide the
correlations between the elements.

We have demonstrated the applicability of the suggested theory to
the different kinds of correlated stochastic systems. However,
there exist some aspects which cannot be interpreted in terms of
our two-parameter model. Obviously, more complex models should be
developed for the adequate description of real correlated systems.

We acknowledge to Dr. S.V. Denisov who drew our attention to the
exposed problem, Yu.L. Rybalko for consultations and kind
assistance in the numerical simulations, S.S. Mel'nik and M.E.
Serbin for the helpful discussions.

\end{document}